\documentclass[a4paper,12pt,oneside,article]{memoir}

\setlrmarginsandblock{3cm}{*}{1}
\setulmarginsandblock{3cm}{*}{1}
\setheadfoot{2cm}{\footskip}         
\checkandfixthelayout[nearest]

\usepackage{hyperref}
\usepackage[utf8]{inputenc} 
\usepackage[T1]{fontenc} 
\usepackage{lmodern}
\usepackage{amsmath, amssymb, bm, mathtools, mathdots}
\usepackage{lineno}

\usepackage[normalem]{ulem}
\usepackage{array, booktabs, tabularx}
\usepackage{graphicx, caption, subfig, xcolor}
\usepackage[retainorgcmds]{IEEEtrantools}
\usepackage{enumitem, lipsum}
\usepackage{cancel}
\usepackage{cite}
\usepackage{authblk}

\graphicspath{ {./images/} }
\usepackage{enumerate}
\def\rm{\mathrm}
\def\Km{\;\rm{km}}
\def\mm{\;\rm{mm}}

\def\m{\;\rm{m}}
\def\km{\;\rm{km}}
\def\bn{\begin{enumerate}}
\def\en{\end{enumerate}}
\def\be{\begin{equation}}
\def\ee{\end{equation}}
\def\bea{\begin{eqnarray}}
\def\eea{\end{eqnarray}}
\def\ba{\begin{array}}
\def\ea{\end{array}}
\def\nn{\nonumber}

\def\change{}

\author[1,*]{Florian Dubath}
\author[2,3]{Maria Alice Gasparini}
\affil[1]{\small Department of Astronomy, University of Geneva, Switzerland}
\affil[2]{Institute of Teacher Education, University of Geneva, Switzerland}
\affil[3]{Departement of Physics, University of Geneva, Switzerland}
\affil[*]{E-mail of the corresponding author: florian.dubath@unige.ch}

\begin{document}



\title{Earth’s Radius from a Single Sunrise Image:\\ A Classroom-Ready Activity}



\maketitle

\begin{abstract}
\change{We present a classroom-based activity using a sunrise photograph of Mont Blanc’s shadow taken from Geneva to estimate the Earth’s radius. By determining the direction of solar rays relative to the local vertical and accounting for atmospheric refraction, students can derive an upper-bound approximately 70\% above the accepted value. The discrepancy provides a concrete illustration of modelling assumptions and observational limitations. The activity combines geometric reasoning, basic trigonometry, and order-of-magnitude estimation, allowing students to obtain a physically meaningful result from simple observations. Beyond its quantitative aspects, the approach highlights essential elements of the scientific method, including hypothesis formulation, model construction, uncertainty analysis, and comparison with external data, offering a structured introduction to Nature of Science (NOS) concepts within upper-secondary STEM curricula.}

\end{abstract}

\section*{1. Introduction}

The estimation of Earth’s radius dates back to around 240 BC, when Eratosthenes used the difference in \change{the} Sun's angle at noon \change{on} the summer solstice between Alexandria and Syene (Aswan) to infer Earth's circumference \cite{Era}. \change{Later, in the 11th century, Al-Biruni proposed another method based on observing the horizon from a mountain summit \cite{alBiruni1,alBiruni2}.} Much later, following the French Revolution, Jean-Baptiste Delambre and Pierre Méchain measured part of the meridian between Dunkirk and Barcelona; this measurement did indeed serve as the basis for defining the meter \cite{dela1, dela2}. \\

Both of these historical methods relied on relatively simple means, whereas today the calculation of Earth’s average radius combines gravimetric and astronomical data, requiring more advanced technology \cite{today1, today2}.
All of these approaches, however, involve measurements taken over large distances or at widely separated locations. Under certain conditions, it is also possible to estimate Earth’s radius using only local observations.\\

In this paper, we estimate a conservative upper bound on Earth's radius using a photograph of Mont Blanc’s shadow taken from Geneva, together with the Sun’s elevation at the time the picture was taken.\\

In Section 2, starting from the photograph, we present the observational context and extract the data required for the subsequent analysis, including the camera’s image-plane geometry.
Section 3 derives the direction of the Sun’s rays relative to the local vertical (celestial/solar geometry), which is then used in Section 4 to constrain the Earth’s maximum diameter. In Section 5, this estimate is further corrected by taking into account atmospheric refraction.
Section 6 discusses the results and presents the conclusions.\\

Beyond the quantitative result, this study illustrates a pedagogical approach to scientific inquiry: starting from simple observations, we formulate hypotheses, construct models, make approximations, and validate results. The method emphasizes the role of careful measurement, reasoning, and the iterative refinement of models, providing a concrete example of the scientific method in practice.

\section*{2. Image characteristics and observational data}

\begin{figure}[!h]
	\centering
        	\includegraphics[width=0.8\textwidth]{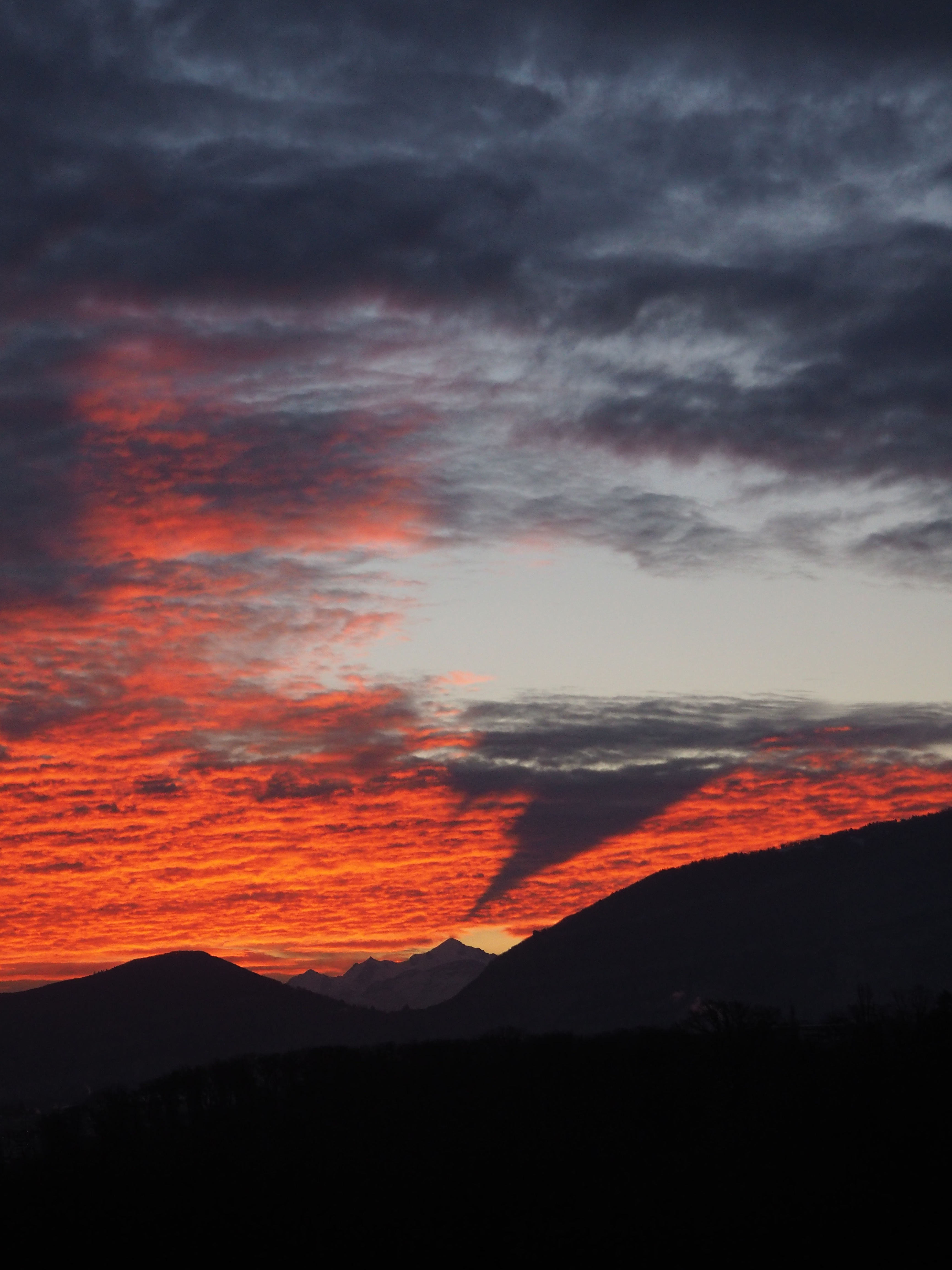}\\
	\caption{The shadows of Mont Blanc and its neighbor, Mont Maudit, projected above their summits onto a cloud layer illuminated by the rising Sun. The sunlight arrives from below the local horizon, as illustrated in the drawing below. 
    }
    \label{fig:photo}
\end{figure}

\begin{figure}[h!]
        	\includegraphics[width=\textwidth]{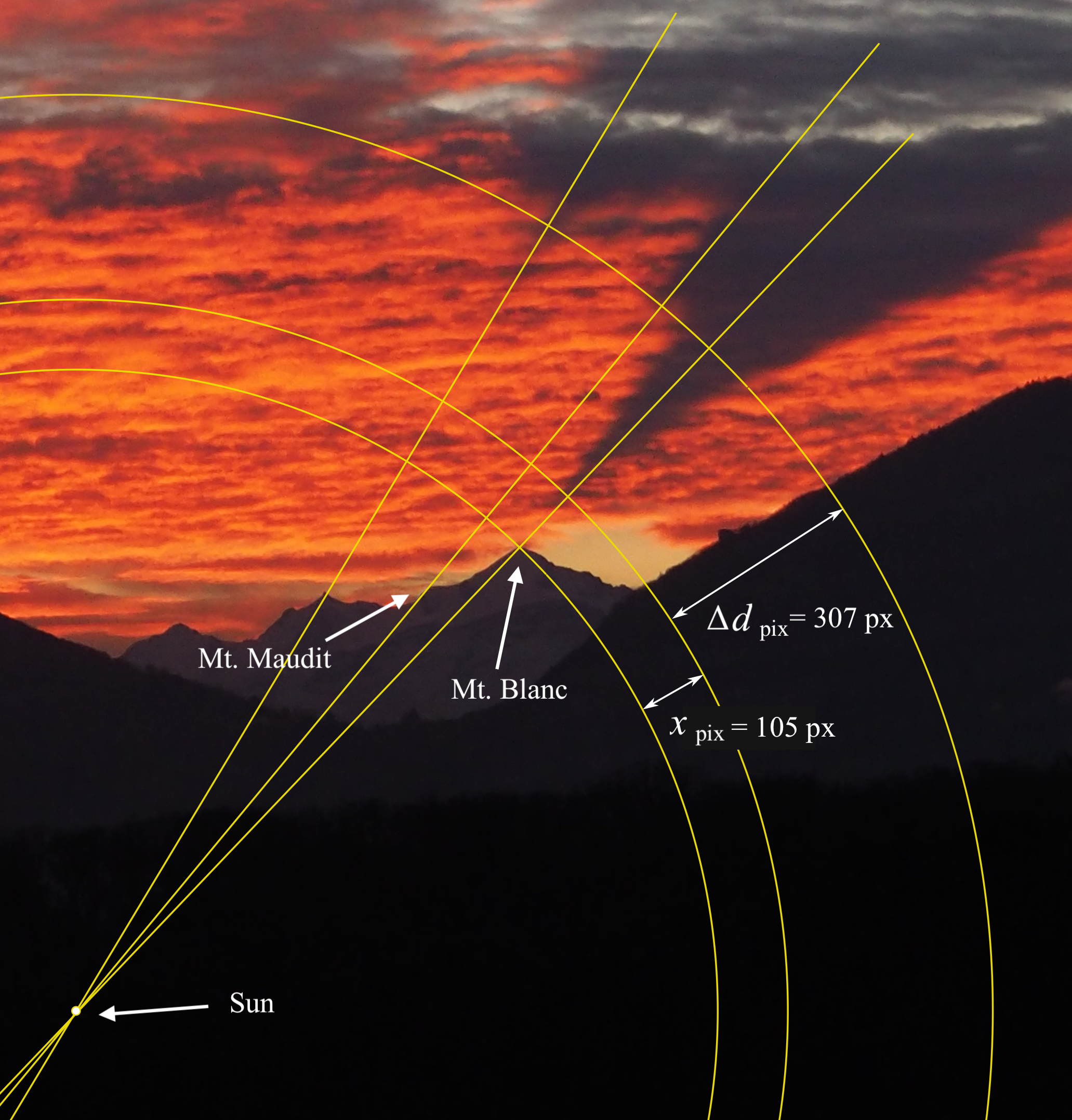}\\
	\caption{\change{Cropped section of the photograph in Fig. 1 showing key details for measurements. Leveraging an additional pair of summit and shadow allows for better localization of the Sun.}}
    \label{fig:cropped}
\end{figure}

The photograph in Fig. \ref{fig:photo} was taken by one of the authors in Geneva on 7 January 2022 at $08:06$ a.m. local time, when the Sun was still below the local horizon of the photographer. The image shows the shadows of Mont Blanc and the adjacent peak Mont Maudit projected onto the overlying stratus cloud layer; the Sun’s position on the photograph corresponds to the intersection of the lines that connect each mountain peak with its respective shadow (see also Fig. \ref{fig:cropped}).

\begin{figure}[!h]
        	\includegraphics[width=\textwidth]{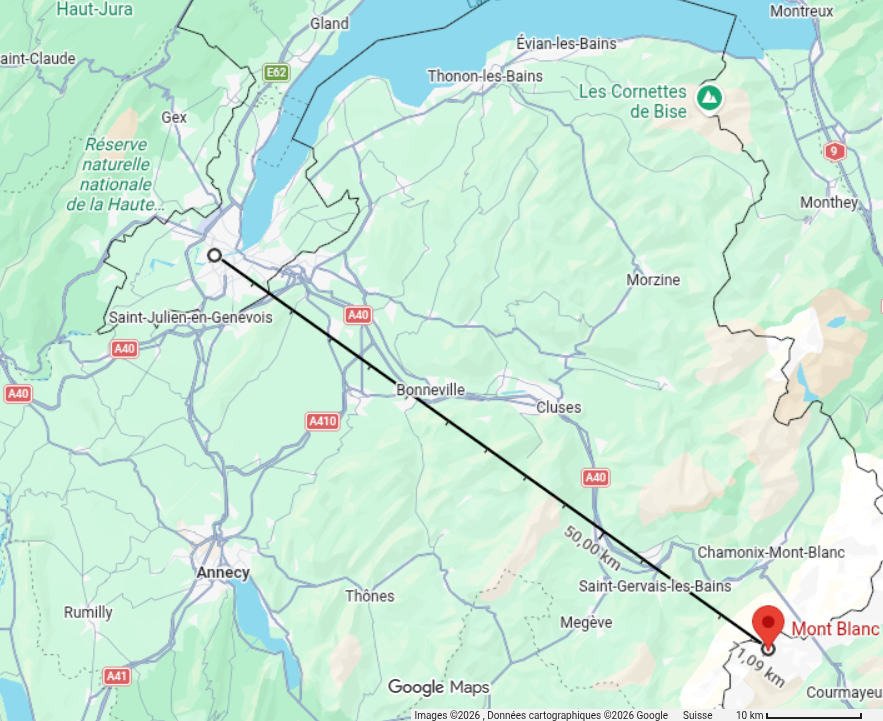}\\
	\caption{\change{Map showing the orientation of the location from which the photo was taken (Geneva) in relation to Mont Blanc (see \url{https://maps.app.goo.gl/JzknKnFRCJMucWiY7})}.
    }
    \label{fig:map}
\end{figure}

As seen on the map in Fig. \ref{fig:map}, this direction is southeast relative to the observer, while the shadows are cast toward the northwest, consistent with the time of sunrise.\\

Mont Blanc and Mont Maudit have elevations of $h_B=4810\m$ and $h_M=4465\m$ above sea level, respectively. The observer is located in Geneva at an altitude of $h_G=430\m$ above sea level. The distance between the photographic location in the Saint-Jean district and the summit of Mont Blanc is $d=71\km$. 

\medskip

Moreover, measurements performed on the cropped photograph, as illustrated in Fig. \ref{fig:cropped}, indicate that the radial pixel separation between the two shadows is $\Delta d_{\rm{pix}}=307$ px, while the total image height of Fig. \ref{fig:photo} is $4608$ px. The radial separation between Mont Blanc and its shadow is $x_{\rm{pix}}=105$ px.\\

\medskip

Although the pixel measurements extracted from the image provide relative distances, converting them into angular separations requires knowledge of the lens’s focal length. The camera used for this photograph is equipped with a $4:3$ sensor ($18.0\mm \times 13.5 \mm$) and a lens of focal length $f = 42.0 \mm$. These data allow us first to convert the measured pixel lengths into physical distances on the sensor. Then, using the elementary optics of a converging lens, we can deduce the corresponding angular separations.\\

Therefore, the radial distances on the \change{camera's sensor} between the two shadow peaks ($\Delta d_{\rm{photo}}$) and between Mont Blanc and its shadow ($x_{\rm{photo}}$) are found to be

\begin{equation}\label{d1phot}
    \Delta d_{\rm{photo}}=\frac{307\;\rm{px}}{4608\;\rm{px} }\cdot 18.0 \;\rm{mm} = 1.20 \;\rm {mm}\quad{\rm{and}}
\end{equation}
\begin{equation}
    x_{\rm{photo}}=\frac{105\;\rm{px}}{4608\;\rm{px}}\cdot 18.0\;\rm{mm} = 0.410\;\rm{mm}\,.
\end{equation}
\medskip

\begin{figure}[!h]
	\centering
        	\includegraphics[width=0.9\textwidth]{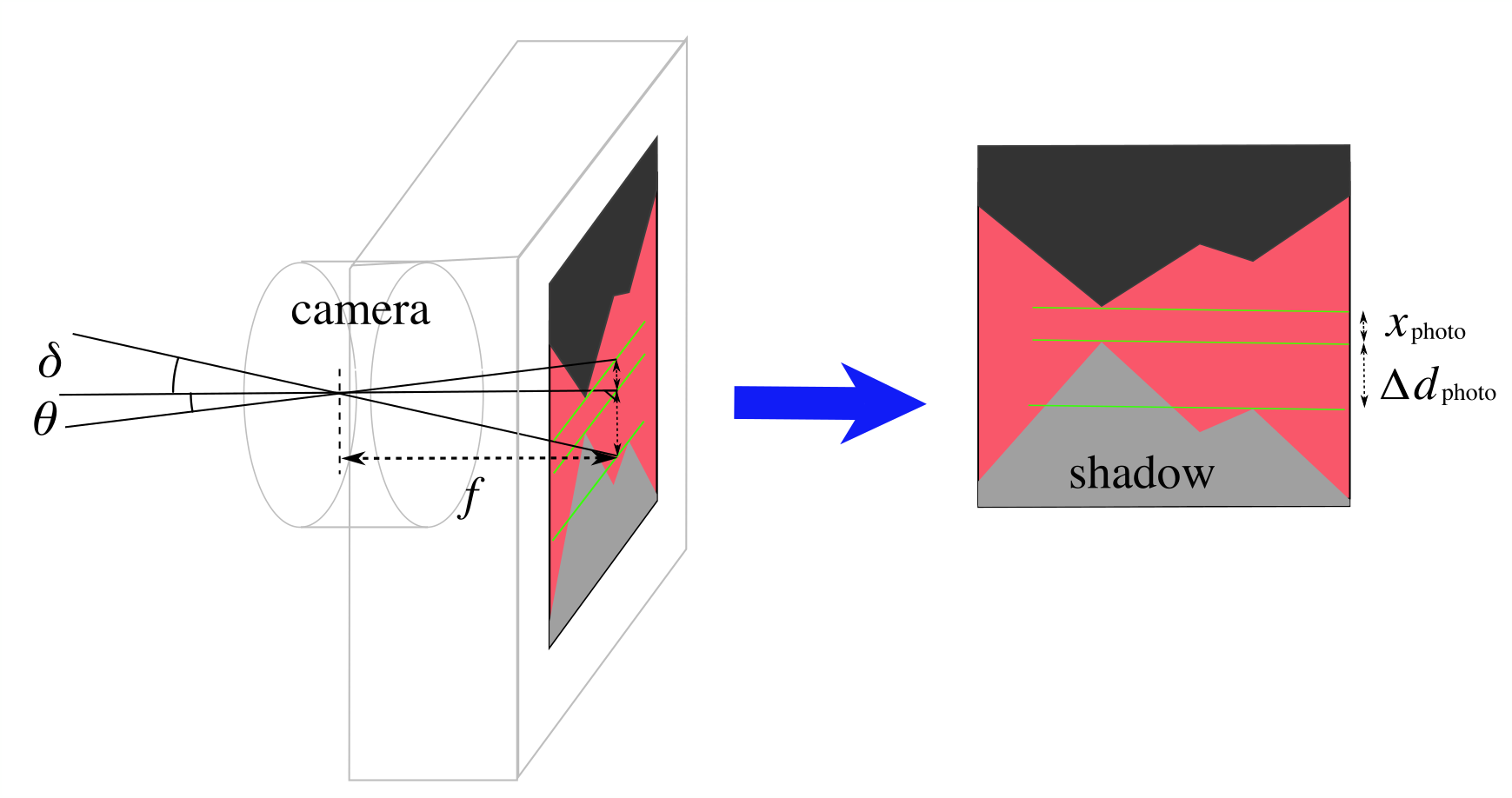}\\
	\caption{\change{Schematic illustrating the image formation on the camera’s sensor.}
    }
    \label{fig:camera}
\end{figure}

Since the distance to the objects (the mountains and their surroundings) relative to the camera is more than five orders of magnitude greater than the distance from the lens to the image, we can reasonably assume that, in the lens equation, the image distance relative to the lens is approximately equal to the focal length. This allows us to determine the corresponding angular separations, denoted $\delta$ and  $\theta$ in the schematic of Fig. \ref{fig:camera}:
\begin{equation}
    \delta = \arcsin\Bigg(\frac{\Delta d_{\rm{photo}}}{f}\Bigg) =  \arcsin\Bigg(\frac{1.20}{42.0}\Bigg) =1.64^\circ \quad{\rm{and}}
\end{equation}
\begin{equation}\label{theta}
    \theta = \arcsin\Bigg(\frac{x_{\rm{photo}}}{f}\Bigg) =  \arcsin\Bigg(\frac{0.41}{42.0}\Bigg) = 0.560^\circ\, .
\end{equation}

\section*{3. Direction of the incident sun rays}

\begin{figure}[!h]
	\centering
        	\includegraphics[width=\textwidth]{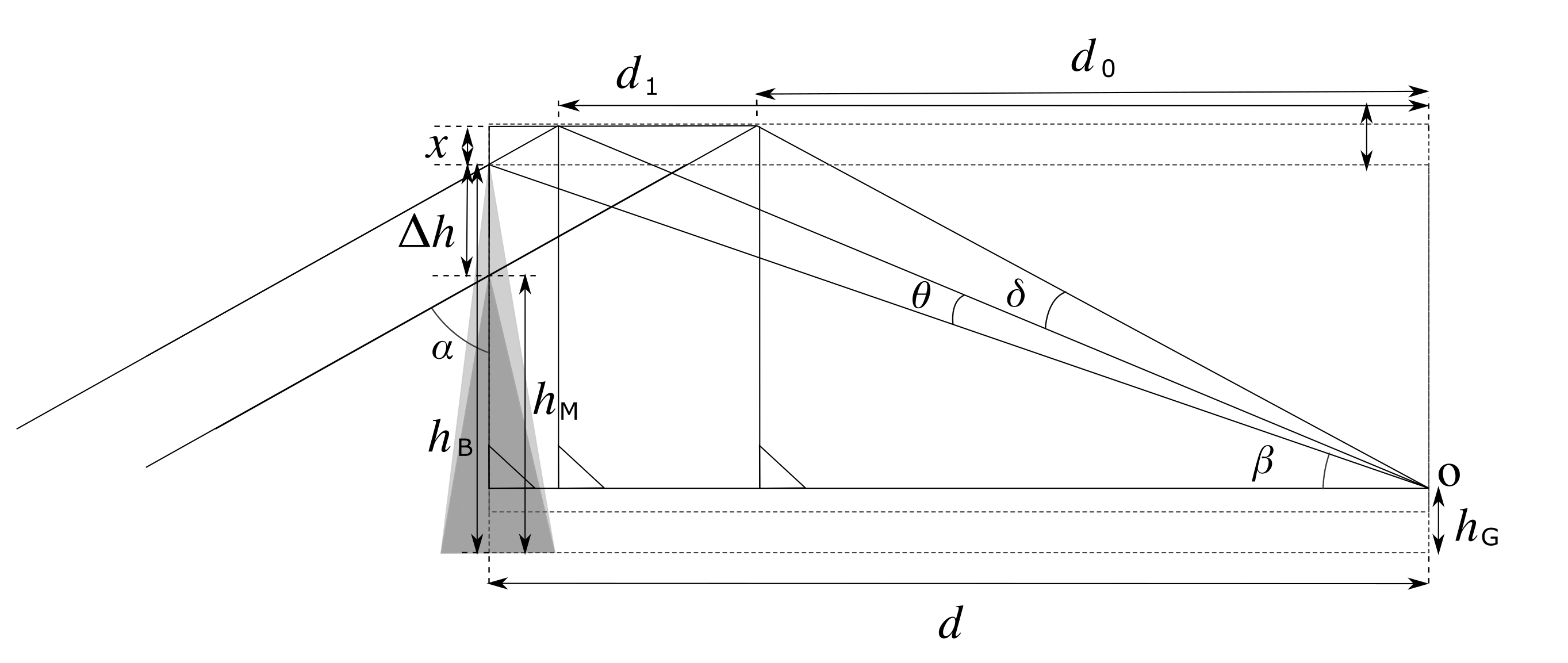}\\
	\caption{Schematic illustration of the situation. The dimensions are not drawn to scale, as the horizontal distance $d$ is more than ten times greater than the height of the cloud layer. 
    }
    \label{fig:schema}
\end{figure}

In this section, we calculate the angle at which the Sun’s rays arrive, relative to the local vertical at the summit of Mont Blanc. We denote this angle as the \textit{angle of light}, $\alpha$, as shown in the diagram in Fig. \ref{fig:schema}.\\

To perform this calculation, we assume that the cloud layer is horizontal and that the curvature of the Earth is negligible over the distance of approximately $\sim 71 \km$. The cloud layer is located above the summit of Mont Blanc at an unknown height.\\

Referring to the schematic in Fig. \ref{fig:schema}, we define the following known distances and angles relevant to the analysis.
\medskip

\begin{itemize}
\setlength\itemsep{0.5em}
    \item The difference in elevation between Mont Blanc ($h_B$) and Mont Maudit ($h_M$) $\Delta h = 345 \m$;
    \item The horizontal distance between the observer and the Mont Blanc summit $d=71\km$; 
    \item The angle between the direction of the summit of Mont Blanc and its shadow, as seen from the observer $\theta = 0.560^\circ$ (Sect. 2);
    \item The angular separation between the shadows of the summits of Mont Blanc and Mont Maudit $\delta = 1,64^\circ$ (Sect. 2);
    \item The angle between the horizontal plane and the direction of the Mont Blanc summit, as seen by the observer:
    \begin{equation}
    \beta =\arctan\Bigg(\frac{h_B - h_G}{d}\Bigg) =  \arctan\Bigg(\frac{4810 - 430}{71\cdot 10^3}\Bigg) =  3.53^\circ\, .
     \end{equation}
\end{itemize} 

The following distances are the unknown variables: 
\begin{itemize}
\setlength\itemsep{0.5em}
    \item The vertical distance between the Mont Blanc summit and the cloud layer $x$;
    \item The horizontal distance between the shadows of the two peaks in the cloud layer $d_1$;
    \item The horizontal distance between the shadow of Mont Maudit and the observer $d_0$.
   
\end{itemize} 

\medskip

From the data presented above and applying right-triangle trigonometry, we can first determine the relation between $x$ and the distances $d_0$ and $d_1$:
\smallskip

\begin{equation}\label{d0}
    d_0 = \frac{x+h_B-h_G}{\tan(\beta + \theta +\delta )}\,;
\end{equation}
\smallskip

\begin{equation}\label{d1}
    d_1 = \frac{x+h_B-h_G}{\tan(\beta + \theta )}\,.
\end{equation}
\smallskip

Next, considering the two similar triangles whose hypotenuses correspond to the parallel light rays, located in the upper-left corner of the diagram, we can write:

\begin{equation}\label{light1}
 \frac{d-d_0}{d-d_1} = \frac{x+\Delta h}{x}\, .
\end{equation}
\medskip

We thus obtain a system of three equations and three unknowns, which we can solve by substituting the expressions for $d_0$ and $d_1$ from Eqs.\eqref{d0} and \eqref{d1} into Eq. \eqref{light1} and introduce the notation $\tan_1 = \tan(\beta + \theta)$, and $\tan_2 = \tan(\beta + \theta +\delta)$ and $h=h_B-h_G$:

\bea
   && \frac{d-(x+h)/\tan_2}{d-(x+h)/\tan_1} = \frac{x+\Delta h}{x}\quad\iff\quad x\left(d-\frac{x+h}{\tan_2}\right) = (x+\Delta h)\left(d-\frac{x+h}{\tan_1}\right)\nn\\
&&\nn\\
&&\nn\\
   &&\iff\quad -\frac{x^2}{\tan_2} + x\left(d - \frac{h}{\tan_2}\right) =  -\frac{x^2}{\tan_1} + x\left(d -\frac{\Delta h + h}{\tan_1}\right) + \Delta h \left(d - \frac{h}{\tan_1}\right) \nn\\
&&\nn\\
&&\nn\\
   &&\iff\quad x^2\left(\frac{1}{\tan_1}-\frac{1}{\tan_2}\right) + x\left(\frac{\Delta h + h}{\tan_1} - \frac{h}{\tan_2}\right)- \Delta h \left(d - \frac{h}{\tan_1}\right) = 0\,.
\eea
\medskip

By replacing the numerical values, this gives
\begin{equation}\label{xeq}
    4.01 x^2 + 2.24 \cdot 10^4 x - 3.36\cdot 10^6 = 0
\end{equation}

Solving Eq. \eqref{xeq} for $x$, we find that the positive solution is $x = 146.17$. This implies that the cloud layer lies approximately $146\m$ above the summit of Mont Blanc, i.e., at an altitude of about $4956\m$ above sea level.\\

As a consistency check, we consulted data from {\textit{MétéoSuisse} and the Geneva Airport weather service corresponding to the time of the observation. Both sources report cloud heights of ``around $5000 \m$'' \cite{meteosuisse}. This value is also consistent with ceilometer measurements above Annecy and Geneva for 7 January 2022 \cite{ceilometer}. 
\\

The angle of light $\alpha$ can therefore be obtained as follows: 
\bea\label{alpha}
   \alpha &=& \arctan{\Bigg(\frac{d-d_0}{x+\Delta h}\Bigg)} = \arctan{\Bigg(\frac{d-h/\tan_2}{x+\Delta h}\Bigg)} \nn\\
   &&\nn\\
   &=& \arctan{\Bigg(\frac{28.0\cdot 10^3  }{496}\Bigg)} = 88.9^\circ
\eea
\bigskip

\subsection*{Angle check}

We obtained the angle of light from the angles $\delta$ and $\theta$ measured directly from the pixel coordinates of the photograph \ref{fig:cropped} (Sect. 2). It is worth noting that a theoretically simpler approach would be to determine, on the same photograph, the angle measured by the observer between the direction of the summit of Mont Blanc and the apparent position of the Sun , i.e., the angle $\gamma$ in Fig. \ref{fig:gamma}. In principle, the Sun’s position can be extrapolated by extending the lines connecting each summit with its corresponding shadow, as illustrated in Fig. \ref{fig:cropped}. \\

If gamma were known with sufficient accuracy, the angle of light $\alpha$ could then be obtained directly using the relationship

\begin{equation}\label{alpha_gamma}
\alpha=\beta + 90^\circ - \gamma\,.
\end{equation}

 \begin{figure}[h!]
	\centering
        	\includegraphics[width=0.8\textwidth]{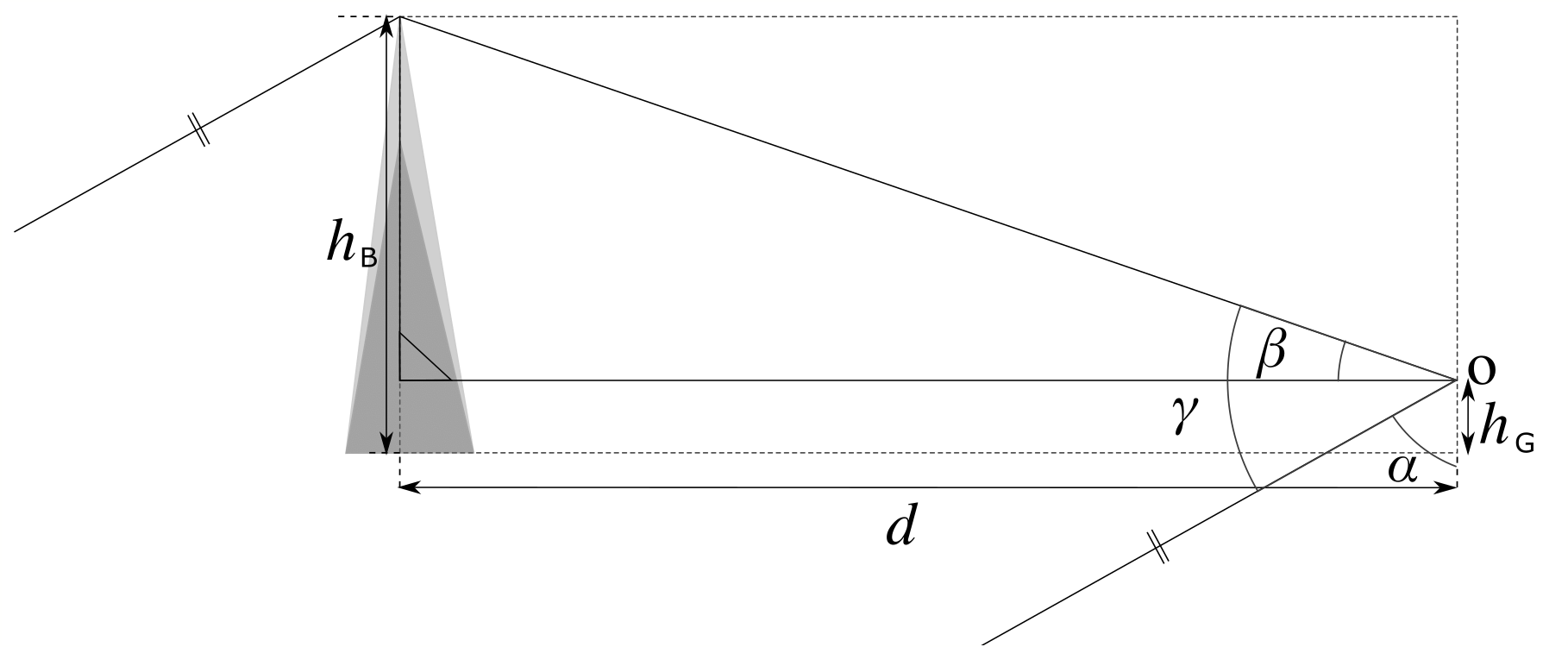}\\
	\caption{Schematic illustrating of the angle observed between the summit of the Mont Blanc and the reconstructed position of the Sun
    }
    \label{fig:gamma}
\end{figure}

However, estimating the Sun’s position on the image \textit{without prior knowledge of} $\alpha$ leads to an uncertainty of approximately $20\%$ in the value of $\gamma$, and the small-angle approximation begins to lose validity. For this reason, we adopted the more elaborate but substantially more precise method based on $\delta$ and $\theta$.\\

That said, we can check that the ``extrapolated'' solar position on the photograph is reasonably consistent with the value of $\alpha$ obtained in Eq. \eqref{alpha}, namely about 866 px above the summit of Mont Blanc.

\begin{equation}\label{alpha_gamma}
\gamma=\beta + 90^\circ - \alpha\quad \Rightarrow\quad \gamma=3.53^\circ + 90^\circ - 88.9^\circ = 4.63^\circ
\end{equation}

\bea
\Rightarrow\quad && D_{\rm{Sun-B\,photo}} \approx f\cdot \sin{\gamma} = 42\mm \cdot \sin{(4.63^\circ)}=3.39\mm\nn\\
&&\nn\\
\Rightarrow\quad && D_{\rm{Sun-B\,pix}} \approx 3,39 \mm \cdot \frac{4608\,\rm{px}}{18\mm} \approx 868\,\rm{px}\,.
\eea

From Fig. \ref{fig:cropped}, the extrapolated position of the Sun is located 961 px from the summit of Mont Blanc. A difference of 100 px corresponds to less than 0.4 mm on the sensor, implying that the graphical precision of this extrapolation is about $10\%$.

\section*{4. Upper bound on Earth radius}

\change{With the angle $\alpha$ determined, the Earth’s radius can be calculated following Al-Biruni’s method \cite{alBiruni2}.}
We assume that Mont Blanc is located on a perfectly spherical Earth and that the light casting its shadow is tangent to the Earth's surface, as shown in Fig. \ref{fig:earth}. In reality, the light may not be perfectly tangent -- either because the photograph was taken too late or because the tangent ray is blocked by other mountains. In such cases, the computed Earth radius would be smaller than our result. Therefore, the following calculation should be considered an upper bound on the Earth's radius.\\

With this in mind, and referring to Fig. \ref{fig:earth}, we can write:

\begin{equation}\label{bondeq}
    \sin(\alpha) = \frac{R}{R+\change{h_B}}
\quad\iff\quad
    R = h_B \cdot \frac{\sin(\alpha)}{1-\sin(\alpha)} 
\end{equation}
\medskip

By substituting the numerical values of $\alpha$, Eq. \eqref{alpha} and $h_B$ into Eq.\eqref{bondeq}, the resulting upper bound on the Earth’s radius is :
\smallskip

\begin{equation}
    R_{\rm{max}} = 4810\cdot \frac{\sin(88.9^\circ)}{1-\sin(88.9^\circ)} = 26.6\cdot 10^3 \; \rm{Km} 
\end{equation}
\smallskip

 \begin{figure}[h!]
	\centering
        	\includegraphics[width=0.8\textwidth]{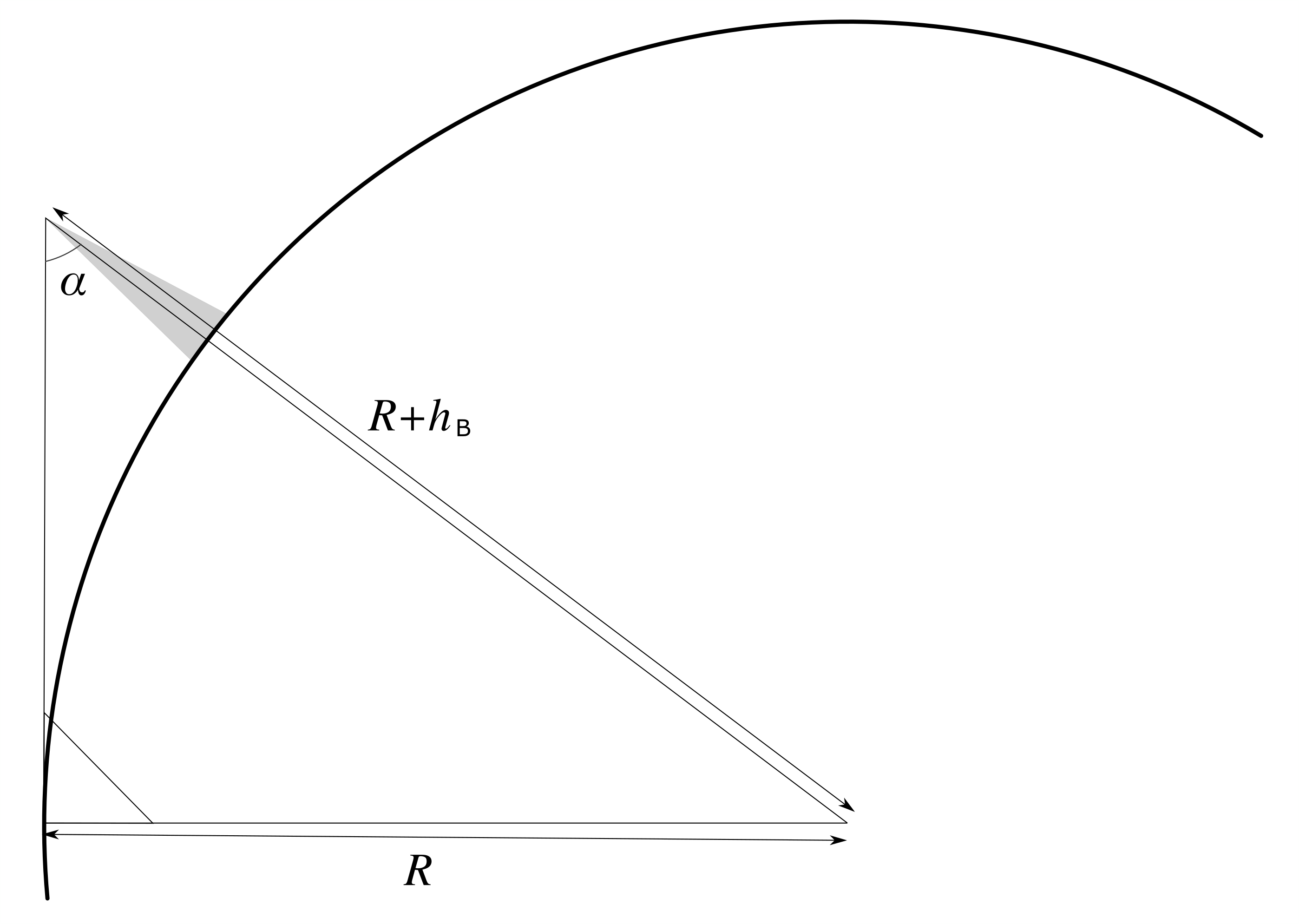}\\
	\caption{Schematic illustrating a tangent ray of sunlight passing the summit of Mont Blanc. The dimensions are not to scale, as the Earth’s radius is more than three orders of magnitude greater than the height of Mont Blanc.  
    }
    \label{fig:earth}
\end{figure}
\medskip
The actual Earth radius is $R_T=6310\Km$ so the bound we found is

\begin{equation}
    R_{\rm{max}} = 4.2 R_T
\end{equation}
\medskip

\section*{5. Refraction}

The estimate given above can be refined by accounting for atmospheric refraction. Since the refractive index of air decreases with altitude, the path of a light ray that is tangent to Earth’s surface and reaches the summit of Mont Blanc acquires a downward curvature, as shown in Fig. \ref{fig:earthrefr}. As a result, for the same angle of light, the observed position of the Sun appears higher than its actual geometric position. Using the standard refraction correction at the horizon, this effect corresponds to an adjustment of \change{$34'$} to the angle $\alpha$ \cite{refraction} \change{(see also \cite{refraction2} and references therein).}\\

 \begin{figure}[h!]\label{earthrefr}
	\centering
        	\includegraphics[width=0.8\textwidth]{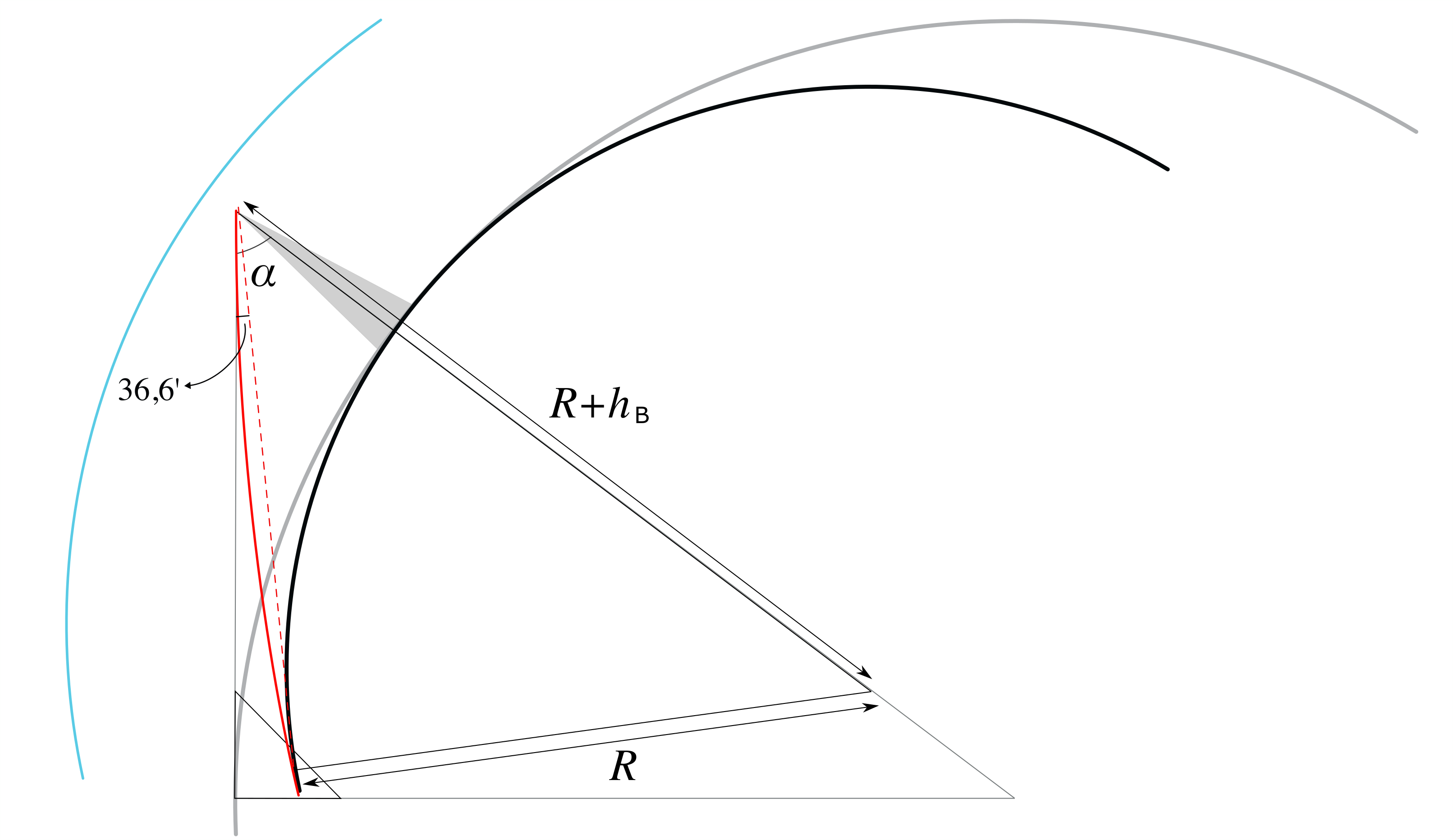}\\
	\caption{Same as Fig. \ref{fig:earth}, with the same angle of light $\alpha$, but here the incoming ray is curved due to atmospheric refraction. This leads to a smaller estimate of the Earth’s radius. 
    }
    \label{fig:earthrefr}
\end{figure}
\medskip

Applying this correction, we obtain:
\begin{equation}
    \alpha_{\rm{corr}} = \alpha - 36.6' = 88.9^\circ-\change{34'} = 88.3^\circ 
\end{equation}

With this value, the bound on the radius becomes 

\begin{equation}
    R_{\rm{max\,corr}} = h_B \frac{\sin(\alpha_{\rm{corr}})}{1-\sin(\alpha_{\rm{corr}})} = 4810\frac{\sin(88.3^\circ)}{1-\sin(88.3^\circ)} = 10.9\cdot 10^3 \Km 
\end{equation}
\medskip

\begin{equation}
   \Rightarrow \quad R_{\rm{max\,corr}} = 1.7 R_T
\end{equation}

Note that the value of the refraction correction at the horizon can vary widely depending on location and meteorological conditions \cite{refractionerror1},\cite{refractionerror2}.

\section*{6. Discussion of the results and concluding remarks}

Using a photograph taken at sunrise from Geneva, showing the shadow of Mont Blanc projected onto the cloud layer above its summit, we estimated the angle of incidence of the solar rays relative to the local vertical (Sect. 3). Assuming a perfectly spherical Earth, we then derived an estimate of the Earth’s radius, first neglecting atmospheric effects (Sect. 4) and subsequently incorporating the deviation of the incident ray caused by the variation of the atmospheric refractive index (Sect. 5).\\

In the derivation in Sect. 4, the available data and procedure allow us to determine the light angle with a precision of three significant digits, corresponding to an uncertainty of roughly 1\%. However, when calculating the Earth’s radius, several factors act to increase this uncertainty, and all of them bias the result towards an overestimate.

\begin{enumerate}
\setlength\itemsep{0.5em}
    \item The first cause of overestimation arises from atmospheric refraction: the light casting Mont Blanc’s shadow passes through the atmosphere, which makes the apparent position of the Sun slightly higher than its true position and therefore yields a higher value of the angle $\alpha$. We accounted for this effect in Sect. 5, which leads to a result closer to the true value of the Earth’s radius.
    \item A second source of uncertainty comes from the fact that the Earth is not a perfect sphere. Local topography -- variations in ground elevation relative to the mean terrestrial radius -- implies that the solar ray producing the shadow may not be tangent to the spherical Earth, but instead lie slightly above it.
    \item The third source of uncertainty arises from Eq. \eqref{bondeq}, which gives the radius as a function of the angle of light. As illustrated in Fig. \ref{fig:bondeq}, even a variation of about 1\% in the estimate of $\alpha$ leads to a change of more than 10\% in the inferred Earth radius.

    This sensitivity is consistent with what we expect from the limiting cases. On the one hand, if the radius $R$ tends toward zero, the incident light rays become nearly vertical with respect to Mont Blanc, and $\alpha\rightarrow 0^\circ$. On the other hand, if $R$ tends toward infinity (flat-Earth limit), the angle of light becomes nearly horizontal, so $\alpha\rightarrow 90^\circ$. 
\end{enumerate}

 \begin{figure}[!h]
	\centering
        	\includegraphics[width=0.7\textwidth]{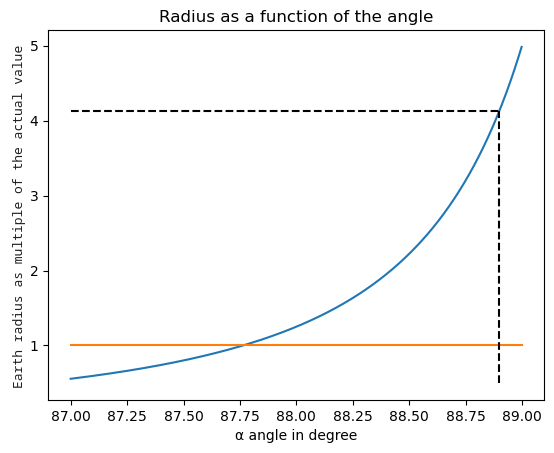}\\
	\caption{
     \change{Graph showing the shape of the function \eqref{bondeq}, which gives the estimate of the Earth's radius $R$ as a function of the angle of light. The value found in Sect. 4 is indicated.} 
    }
    \label{fig:bondeq}
\end{figure}
\medskip

Beyond the numerical precision achieved, this work has pedagogical value from several perspectives. It shows how simple and careful observations of an everyday situation -- one that might initially appear mundane -- can be used to construct key steps in the scientific method and raise awareness of the nature of science (NOS).

 This activity allows students to formulate relevant qualitative questions, which can then be translated into quantitative estimates through geometric and mathematical modelling. It develops the ability to make and account for approximations during problem-solving, compare alternative approaches, evaluate uncertainties, and study limiting cases as a method for verifying the relevance of results.

Moreover, the modest technical level required -- no more than upper-secondary school -- and the authenticity and serendipity of the situation make this work a valuable example of a pedagogical activity capable of stimulating students’ motivation and interest in physics classes.

\listoffigures

\end{document}